\documentclass[12pt]{iopart}
\usepackage[dvips]{graphicx}
\begin{document}

\newcommand{\beq}{\begin{equation}}
\newcommand{\eeq}{\end{equation}}
\newcommand{\barr}{\begin{eqnarray}}
\newcommand{\earr}{\end{eqnarray}}

\newcommand{\andy}[1]{ }

%%%%%%%%%%%%%%%%%%%%%%%%new def%%%%%%%%%%%%%%%%%%%%%%%%%%%
\def\bra#1{\langle #1 |}
\def\ket#1{| #1 \rangle}
\def\cH{{\cal H}}
\def\cG{{\cal G}}
\def\cU{{\cal U}}
\def\bmp{\mbox{\boldmath $p$}}
\renewcommand{\Im}{\mathrm{Im}}
%%%%%%%%%%%%%%%%%%%%%%%%%%%%%%%%%%%%%%%%%%%%%%%%%%%%%%%%%%

\title[Decoherence, fluctuations and Wigner function
in neutron optics]{Decoherence, fluctuations and Wigner function
in neutron optics}

\author{P. Facchi,\(^{1}\) A. Mariano,\(^{1}\)
S. Pascazio,\(^{1}\) and M. Suda\(^{2}\)}

\address{$^{1}$
Dipartimento di Fisica, Universit\`a di Bari
     and Istituto Nazionale di Fisica Nucleare, Sezione di Bari,
 I-70126 Bari, Italy }

\address{$^2$\"Osterreichisches Forschungszentrum, Seibersdorf,
Austria}

\begin{abstract}
We analyze the coherence properties of neutron wave packets, after
they have interacted with a phase shifter undergoing different
kinds of statistical fluctuations. We give a quantitative (and
operational) definition of decoherence and compare it to the
standard deviation of the distribution of the phase shifts. We
find that in some cases the neutron ensemble is more coherent,
even though it has interacted with a wider (i.e.\ more disordered)
distribution of shifts. This feature is independent of the
particular definition of decoherence: this is shown by proposing
and discussing an alternative definition, based on the Wigner
function, that displays a similar behavior. We briefly discuss the
notion of entropy of the shifts and find that, in general, it does
not correspond to that of decoherence of the neutron.

\end{abstract}



%\maketitle

\section{Introduction}

Decoherence is an interesting phenomenon, related to the
long-standing issue of irreversibility. Nowadays, it discloses
challenging perspectives in the light of new technologies and
related physical applications. There is a widespread consensus
\cite{Dec,NPN,Zurek} about the meaning of decoherence, viewed as
the loss of quantum mechanical coherence of a physical system in
interaction with other systems (``environment"). However, a {\em
quantitative} definition of decoherence is subtle and involves
conceptual pitfalls \cite{fibonacci}. In addition, it always
depends on the experimental configuration. An interesting quantity
in this context is the square of the density matrix
\cite{Watanabe}. Apart from lacking idempotency for mixed states,
this quantity enjoys other interesting features \cite{Manfredi},
but also yields results which are at variance with naive
expectations based on entropy \cite{fibonacci}.

In this article we will consider two different definitions of
decoherence: the first is operational and stems from an analysis
of the visibility in quantum (as well as classical, as we will
see) interference experiments. We stress that these experiments
are routinely performed in neutron optics
\cite{neutron,BRSW,postsel}. The second definition is based on the
idempotency defect of the density matrix and is, in this sense,
less operational.

In both cases, decoherence displays an ``anomalous'' behavior,
both as a function of the features of  the fluctuations and the
incoming state. Some concrete examples will be considered and
discussed. Our analysis will focus on neutron optics and hinge on
an approach based on the analysis of statistical fluctuations
\cite{sim,RS}. However, since our results are just a
consequence of the wave nature of neutrons and their coherence
properties, we expect that the same general conclusions be valid
for other quantum (and classical) waves.

\section{Fluctuations in neutron optics}
\label{sec-nof}\andy{sec-nof}

\begin{figure}[t]
\begin{center}
\begin{tabular}{c}
\includegraphics[width=12cm]{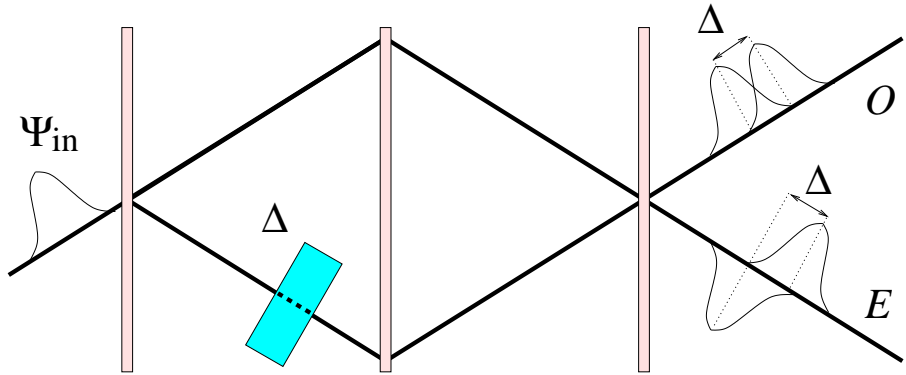}
\end{tabular}
\end{center}
\caption{Scheme of a Mach-Zehnder interferometer.}
\label{fig:inter}
\end{figure}

Let us start our analysis by considering a neutron beam that
crosses a Mach-Zehnder interferometer (MZI), as schematically
shown in Fig.\ \ref{fig:inter}. A phase shifter $\Delta$ is placed
in the lower arm of the interferometer and $\ket{\psi_{\rm in}}$
is the initial wave packet.

We neglect wave-packet dispersion effects, so that the outgoing
states in the ordinary and extraordinary channels read
\andy{uscint-}
\barr
\ket{\psi_{O}} &=&\frac{1}{2}\left[1+ e^{\frac{i}{\hbar}\hat
p\Delta}\right]\ket{\psi_{\rm in}}, \nonumber  \\
\label{eq:uscint-}
\ket{\psi_{E}}&=&\frac{1}{2}\left[1- e^{\frac{i}{\hbar}\hat
p\Delta}\right]\ket{\psi_{\rm in}},
\earr
respectively. We focus on the ordinary channel, the analysis for
the extraordinary one being identical. Define the operator
\andy{Odef}
\beq\label{eq:Odef}
\hat
O(\Delta) = \frac{1}{2}\left[1+ e^{\frac{i}{\hbar}\hat
p\Delta}\right],
\eeq
that accounts for the state evolution in the ordinary channel, and
consider the output density matrix
\andy{matde}
\beq
\label{eq:matde}
\rho_{O}\equiv\ket{\psi_{ O }}\bra{\psi_{O }}=\hat
O(\Delta)\ket{\psi_{\rm in}}\bra{\psi_{\rm in}} \hat
O(\Delta)^\dagger=\hat O(\Delta)\rho_{\rm in}\hat
O(\Delta)^\dagger,
\eeq
where $\rho_{\rm in}$ is the density matrix of the incoming state.
The trace of $\rho_{O}$ yields the relative frequency of neutrons
in the ordinary channel.

Suppose now that the phase shift $\Delta$ fluctuates according to
a probability law $w(\Delta-\Delta_0), \; \Delta_0$ being the
average phase (operationally defined as the phase that is
measured---or inferred \cite{poissongauss}---in an interferometric
experiment). Therefore one has
\andy{wrules}
\beq\label{eq:wrules}
\int d\Delta\; w(\Delta) =1, \qquad
\int d\Delta\; w(\Delta)\Delta =0.
\eeq
The trace of the average density matrix is
\andy{tmadem}
\beq\label{eq:tmadem}
{\rm Tr }\;\overline{\rho_{O}}={\rm Tr
}\;\int\;d\Delta\;w(\Delta-\Delta_0) \hat O(\Delta) \rho_{\rm
in}\hat O(\Delta)^\dagger={\rm Tr }\;\left( \rho_{\rm
in}\overline{\hat O(\Delta)^\dagger\hat O(\Delta)}\right),
\eeq
where the bar denotes the average over the distribution
$w(\Delta-\Delta_0)$. One obtains, after some algebra,
\andy{oocro}
\beq\label{eq:oocro}
\overline{\hat O(\Delta)^\dagger\hat
O(\Delta)}=\frac{1}{2}\left(1+\overline{\cos\frac{\hat
p\Delta}{\hbar}}\right).
\eeq
Consider now the Fourier transform of the probability density of
the fluctuations
\andy{wtilde}
\barr\label{eq:wtilde}
\Omega( p)&\equiv&\int\;d\Delta\; w(\Delta) e^{\frac{i}{\hbar}
p\Delta}
\nonumber \\
&=&\int\;d\Delta\;w(\Delta)\cos\frac{
p\Delta}{\hbar}+i\int\;d\Delta\; w(\Delta)\sin\frac{
p\Delta}{\hbar}\nonumber \\
&=&C( p)+i S( p),
\earr
where $C$ and
$S$ are respectively the real and the imaginary part of $\Omega$
\andy{reome,imome}
\barr \label{eq:reome}
C( p)&=& {\rm Re }\;\Omega ( p), \\
S( p)&=& {\rm Im }\;\Omega( p).
\label{eq:imome}
\earr
In Eq. (\ref{eq:oocro}) we can write
\andy{copdel}
\barr\label{eq:copdel}
\overline{\cos\frac{\hat p \Delta}{\hbar}
}=\int\;d\Delta\;w(\Delta-\Delta_0)\cos\frac{\hat p \Delta }{\hbar
} =\cos\frac{\hat p\Delta_0 }{\hbar }\;C(\hat p)-\sin\frac{\hat
p\Delta_0 }{\hbar }\;S(\hat p).
\earr
In this paper, for simplicity, we will always consider symmetric
distribution functions, that is $w(\Delta)=w(-\Delta)$. Therefore
\andy{evenome}
\beq\label{eq:evenome}
S(\hat p)=0, \qquad C(\hat p)=\Omega(\hat p)
\eeq
and (\ref{eq:oocro}) becomes
\andy{omoco}
\beq\label{eq:omoco}
\overline{\hat O(\Delta)^\dagger\hat
O(\Delta)}=\frac{1}{2}\left[1+\Omega(\hat p)\cos\frac{\hat
p\Delta_0 }{\hbar }\right].
\eeq
We notice, incidentally, that the same results are obtained with a
different setup \cite{BRSW}: consider a polarized neutron that
interacts with a magnetic field perpendicular to its spin. Due to
the longitudinal Stern-Gerlach effect \cite{longSG}, its wave
packet is split into two components that travel with different
speeds and are therefore separated in space. After a projection
onto the initial spin state, the resulting final state is slightly
different from that considered in the preceding equations: we need
to replace $\ket{\psi_{O}}$ (and analogously $\ket{\psi_{E}}$) in
(\ref{eq:uscint-}) with
\andy{magncha}
\beq
\label{eq:magncha}
\ket{\psi_{O}}\longrightarrow \ket{\psi_{O}'}=\hat
O'(\Delta)\ket{\psi_{\rm in}},
\eeq
where
\andy{finmag}
\beq
\label{eq:finmag}
\hat O'(\Delta)=\frac{1}{2}\left[ e^{-\frac{i}{2\hbar}\hat
p\Delta}+e^{\frac{i}{2\hbar}\hat p\Delta}\right],
\eeq
and $\Delta$ is in this case the spatial separation between the
two wave packets corresponding to the two spin components. By
averaging over $\Delta$ it is easy to show that one obtains again
(\ref{eq:omoco}).

By plugging the average operator (\ref{eq:omoco}) into
(\ref{eq:tmadem}) one finally gets
\andy{fintr}
\beq\label{eq:fintr}
{\rm Tr }\;\overline{\rho_O
}=\frac{1}{2}\left[1+\left\langle\Omega(\hat p)\cos\frac{\hat
p\Delta_0 }{\hbar }\right\rangle\right],
\eeq
where $\langle\cdots\rangle = \mathrm{Tr}
\left[\rho_{\mathrm{in}}\cdots\right]$ denotes the expectation
value over the initial state $\rho_{\rm in}$. On the other hand,
the momentum distribution is easily shown to be
\andy{imptr}
\beq\label{eq:imptr}
P_O(p)=\bra p\overline{\rho_O }\ket p =\mathrm{Tr}
\left(\ket{p}\bra{p}\;\overline{\rho_O }\right)=\frac{1}{2}P_{\rm
in}(p) \left[1+\Omega(p)\cos\frac{p\Delta_0}{\hbar }\right], \eeq
where
\andy{imptr0}
\beq\label{eq:imptr0}
P_{\rm in}(p)=\bra p \rho_{\rm in}\ket p.
\eeq
We now introduce the {\em visibility} of the interference pattern
(in the ordinary channel)
\andy{visibp}
\beq
\label{eq:visibp}
{\cal
V}(p)\equiv\frac{P_O(p)_{\rm MAX }-P_O(p)_{\rm min} }{P_O(p)_{\rm
MAX }+P_O(p)_{\rm min} }=|\Omega(p)|,
\eeq
where $P_O(p)_{\rm MAX}$ [$P_O(p)_{\rm min}$] is the maximum
[minimum] value assumed by $P_O(p)$ when $\Delta_0$ varies. By the
very definition (\ref{eq:wtilde}), one can verify that $0\leq
{\cal V}(p)\leq 1$. Notice that, according to this definition, the
visibility is a function of momentum $p$ and yields a measure of
the fringe visibility of a postselected beam of momentum $p$ as a
function of the phase shift $\Delta_0$
\cite{post,BRSW}. Equivalently, it is a measure of the ``local''
spectral visibility, under the assumption of a slowly varying wave
envelope, and so it corresponds to (the absolute value of) the
amplitude of the cosine function in (\ref{eq:imptr}). By using
(\ref{eq:wtilde}) and (\ref{eq:visibp}), one infers that the
visibility is the modulus of the Fourier transform of the
distribution of the shifts $\Delta$ and is therefore a quantity
that is closely related to the physical features of the phase
shifter. In this way we can easily relate the visibility of the
interference pattern (and, as we will see below, the decoherence)
to the ``environmental'' fluctuations. Note that a completely
equivalent definition of the spectral visibility
(\ref{eq:visibp}), which is nevertheless more symmetric and makes
use also of the extraordinary channel, reads
\andy{visibp1}
\beq
\label{eq:visibp1}
{\cal V}(p)= \max_{\Delta_0 }\frac{\left | P_O(p)-P_E(p)
\right|}{P_O(p)+P_E(p)}=\max_{\Delta_0 }\left|
\Omega(p)\cos\frac{p\Delta_0 }{\hbar }\right| = |\Omega(p)|,
\eeq
where the momentum distribution of the extraordinary channel is
given by
\beq
P_E(p)=\frac{1}{2}P_{\rm in}(p)
\left[1-\Omega(p)\cos\frac{p\Delta_0}{\hbar }\right],
\eeq
whence $P_O(p)+P_E(p)=P_{\mathrm{in}}(p)$. The spectral visibility
in the form (\ref{eq:visibp1}) leads to a straightforward
generalization which is at the basis of an operational definition
of decoherence.

\section{An operational definition of decoherence}
\label{sec-cont} \andy{sec-cont}

Let us endeavor to give a quantitative definition of decoherence
based on the definition of visibility given in the previous
section. We start from the relative frequency of particles
detected in the ordinary and extraordinary channels
\andy{OEpar}
\barr
{\cal N}_O(\Delta_0) &=&{\rm Tr }\;\overline{\rho_O }
=\frac{1}{2}\left[1+\left\langle\Omega(\hat p)\cos\frac{\hat
p\Delta_0}{\hbar}\right\rangle\right], \nonumber\\
 {\cal N}_E(\Delta_0) &=&{\rm Tr }\;\overline{\rho_E}
=\frac{1}{2}\left[1-\left\langle\Omega(\hat p)\cos\frac{\hat
p\Delta_0}{\hbar}\right\rangle\right].
\label{eq:OEpar}
\earr
Their difference is
\andy{de0vis}
\beq\label{eq:de0vis}
{\cal N}_O(\Delta_0)-{\cal N}_E(\Delta_0)= \left\langle\Omega(\hat
p)\cos\frac{\hat p\Delta_0}{\hbar}\right\rangle
\eeq
and one can define a {\em generalized visibility}
\andy{de0visb}
\barr
{\cal V}&=&\max_{\Delta_0 }|{\cal N}_O(\Delta_0)-{\cal
N}_E(\Delta_0)| =\max_{\Delta_0 }\left|\left\langle \Omega(\hat
p)\cos\frac{\hat p\Delta_0 }{\hbar }\right\rangle\right|
\nonumber\\
&=&\max_{\Delta_0 }\left|\int dp\;P_{\rm
in}(p)\Omega(p)\cos\frac{p\Delta_0 }{\hbar }\right|.
\label{eq:de0visb}
\earr
It is apparent that (\ref{eq:de0visb}) is the straightforward
generalization of the spectral visibility (\ref{eq:visibp1}),
because obviously ${\cal N}_O+{\cal N}_E=1$. It represents a
\emph{global} feature of the outgoing state, in contrast with the
local character of (\ref{eq:visibp1}). Notice, however, that when
$P_{\rm in}(p')=
\delta(p'-p)$ (incoming monochromatic beam of momentum $p$), the
generalized visibility (\ref{eq:de0visb}) reduces to the standard
``local" visibility (\ref{eq:visibp})
\andy{qwe}
\barr
{\cal V } &=& \max_{\Delta_0}
\left|\int\;dp'\;\delta(p'-p)\Omega(p')\cos\frac{p'\Delta_0}{\hbar
} \right| \nonumber \\
&=& \max_{\Delta_0}
\left|\Omega(p)\;\cos\frac{p\Delta_0}{\hbar}\right|={\cal V}(p)  .
\label{eq:qwe}
\earr
This is a consistency check, because a spectral postselection is
equivalent to injecting an incoming monochromatic beam.

In general one gets
\andy{rty}
\beq\label{eq:rty}
{\cal V}\leq\max_{\Delta_0}\int\;dp\;P_{\rm in}(p)|\Omega(p)|
\left|\cos\frac{p\Delta_0}{\hbar}\right|=\int\;dp\;P_{\rm
in}(p){\cal V}(p)=\left\langle \mathcal{V}(\hat{p}) \right\rangle.
\eeq
The generalized visibility yields the maximum ``distance" between
the intensities ${\cal N}_O$ and ${\cal N}_E$ and is bounded by
the ``local" visibility averaged over the momentum distribution of
the incoming state.

Notice that $P_{\rm in}(p)$ is a nonnegative quantity, while
$\Omega(p)$, being a Fourier transform, is not. For this reason,
in general, the $\max_{\Delta_0}$ does not simply enter the
integral in (\ref{eq:de0visb}), so that (\ref{eq:rty}) is a strict
inequality. However, in the particular case $\Omega(p)
\geq 0$, ${\cal V}$ saturates its upper bound, the equality sign
holds in (\ref{eq:rty}), and Eq.\ (\ref{eq:de0visb}) simplifies
into
\andy{de0visbsimpl}
\barr
{\cal V}=\int dp\;P_{\rm in}(p)\Omega(p). \qquad (\Omega(p)\geq 0)
\label{eq:de0visbsimpl}
\earr
As is often to be expected, the most interesting cases are those
situations in which $\Omega(p)$ is {\em not} always positive,
giving rise to ``anomalous" situations.

\begin{figure}[t]
\begin{center}
\begin{tabular}{c}
\includegraphics[width=9.5cm]{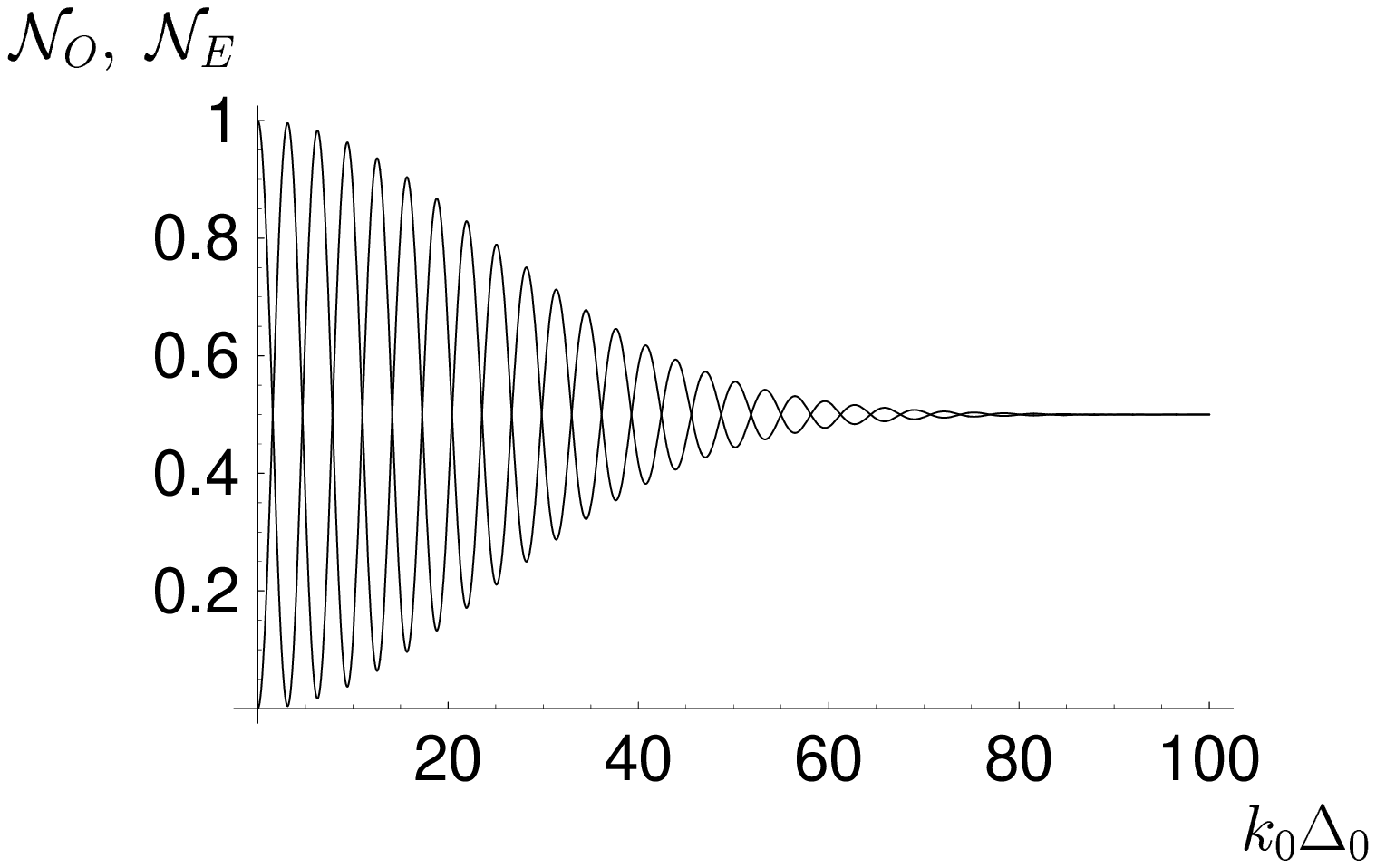}
\end{tabular}
\end{center}
\caption{Relative frequencies of neutrons detected in
the ordinary ${\cal N}_O$ and extraordinary ${\cal N}_E$ channel
versus $k_0\Delta_0$ ($k_0=p_0/\hbar$), for an incoming Gaussian
wave packet (\ref{eq:rum5}) with $k_0\delta=12$ and a
fluctuation-free phase shifter. The two intensities differ in
phase by $\pi$ and their sum is 1. The generalized visibility
(\ref{eq:de0visb}) is 1.}
\label{fig:visi}
\end{figure}

In order to understand the physical meaning of the generalized
visibility, it is useful to look at the example of a
fluctuation-free phase shifter, $w(\Delta)=\delta(\Delta)$, for
which (\ref{eq:wtilde}) yields $\Omega(p)=1$, so that the
generalized visibility (\ref{eq:de0visb}) becomes
\andy{nofluct}
\beq
\label{eq:nofluct}
{\cal V}= \max_{\Delta_0 }\left|\int\;dp\;P_{\rm in}(p)
\cos\frac{p\Delta_0}{\hbar }\right| = \int\;dp\;P_{\rm in}(p)=1,
\eeq
for any incoming distribution $P_{\rm in}$. This result follows
also directly from (\ref{eq:de0visbsimpl}).  For instance, for an
incoming Gaussian wave packet
\andy{rum5}
\beq\label{eq:rum5}
P_{\rm in}(p)=\sqrt{\frac{2\delta^2}{\hbar^2\pi}
}\exp\left(-\frac{2\delta^2}{\hbar^2}(p-p_0)^2\right),
\eeq
one gets the interference patterns ${\cal N}_O$ and ${\cal N}_E$
shown in Figure
\ref{fig:visi}, where it is apparent that ${\cal V}=1$.

If, on the other hand, the phase shifter fluctuates, the amplitude
of the envelope function decreases and ${\cal V}<1$. We therefore
give an operational definition of decoherence, by defining a {\em
decoherence parameter}:
\andy{decimp}
\barr
\varepsilon \equiv 1-{\cal
V}&=&1-\max_{\Delta_0}\left|\left\langle\Omega(\hat
p)\cos\frac{\hat p\Delta_0 }{\hbar }\right\rangle\right|
\nonumber\\
&=& 1 -\max_{\Delta_0 }\left|\int dp\;P_{\rm
in}(p)\Omega(p)\cos\frac{p\Delta_0 }{\hbar }\right| .
\label{eq:decimp}
\earr
Notice that, by Eq.\ (\ref{eq:nofluct}), $\varepsilon=0$ for a
fluctuation-free phase shifter (quantum coherence perfectly
preserved), while $\varepsilon \to 1$ when the magnitude of the
fluctuations increases, $\Omega( p)\to0$ and the envelope function
in Figure \ref{fig:visi} squeezes away all oscillations,
eventually yielding ${\cal N}_O (\Delta_0)={\cal N}_E(\Delta_0)$,
independently of $\Delta_0$. Observe also that ${\cal V}$ and
$\varepsilon$ are independent of the coherence of the initial
state (namely, they do not depend on the off-diagonal terms of the
density matrix). On the other hand, they strongly depend on the
momentum distribution of the initial state (\ref{eq:imptr0}). In
this sense they measure the {\em loss} of quantum coherence caused
by a given physical setup, independently of the coherence of the
incoming state.

It is important to stress that the above definition of decoherence
is {\em operational}. One first measures the relative frequencies
of neutrons detected in the ordinary and extraordinary channels as
a function of $\Delta_0$, both being measurable quantities. Then
one evaluates (\ref{eq:de0visb}) and computes $\varepsilon$.

\section{Some examples}
\andy{sec-capa} \label{sec-capa}

The decoherence parameter (\ref{eq:decimp}) depends on the product
of the momentum distribution of the incoming beam times the
spectrum of the phase-shifter fluctuations, $P_{\rm
in}(p)\times\Omega(p)$. These two ingredients affect $\varepsilon$
at the same level. Therefore, their role can be interchanged: by
maintaining their product unaltered, there exist ``dual"
situations that give exactly the same decoherence parameter with
very different kinds of statistical fluctuations and incoming
states.

\begin{figure}[t]
\begin{center}
\begin{tabular}{c}
\includegraphics[width=8.5cm]{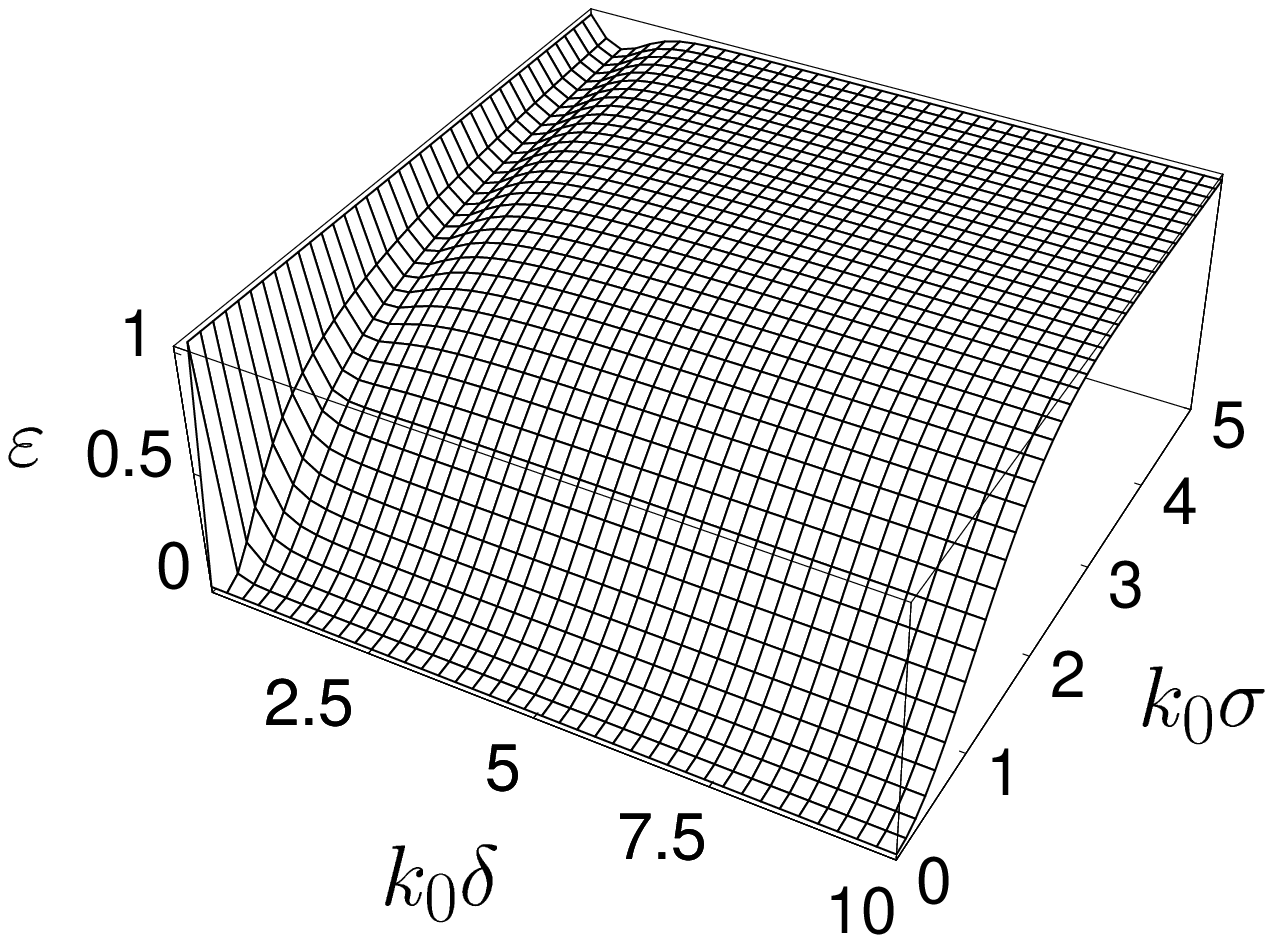}
\end{tabular}
\end{center}
\caption{Decoherence parameter $\varepsilon$ (\ref{eq:rum6})
versus width $\delta$ of the Gaussian wave packet and standard
deviation $\sigma$ of the fluctuating shifts ($k_0=p_0/\hbar$).}
\label{fig:gaun}
\end{figure}

By keeping the above remark in mind,  it is interesting to look at
some particular cases that can be treated analytically. Let the
phases be distributed according to a Gaussian law with standard
deviation $\sigma$
\andy{rum1}
\beq\label{eq:rum1}
w(\Delta-\Delta_0)=\frac{1}{\sqrt{2\pi\sigma^2}
}\exp\left(-\frac{(\Delta-\Delta_0)^2 }{2\sigma^2 }\right),
\eeq
so that $\Omega(p)=\exp\left(-p^2\sigma^2/2\hbar^2\right)$ and the
decoherence parameter reads
\andy{rum3}
\barr
\varepsilon&=&1-\max_{\Delta_0 }\left|\int\;dp\;P_{\rm in}(p)
\exp\left(-\frac{p^2\sigma^2}{2\hbar^2}\right)\cos
\left(\frac{p\Delta_0}{\hbar}\right)\right| \nonumber \\
& = & 1 - \int\;dp\;P_{\rm in}(p)
\exp\left(-\frac{p^2\sigma^2}{2\hbar^2}\right) .
\label{eq:rum3}
\earr
For the Gaussian wave packet (\ref{eq:rum5}) one gets
\andy{rum6}
\beq\label{eq:rum6}
\varepsilon=1-\sqrt{\frac{\delta^2}{\delta^2+\sigma^2/4}
}\exp\left(-\frac{\delta^2}
{\delta^2+\sigma^2/4}\frac{\sigma^2k_0^2}{2}\right),
\eeq
with $k_0=p_0/\hbar$. This is exact and is shown in Figure
\ref{fig:gaun}.
At fixed $\delta$ the decoherence parameter (\ref{eq:rum6})
increases with $\sigma$, although the details of its behavior are
strongly dependent on the spatial width of the packet $\delta$.
This behavior is in agreement with expectation: decoherence
$\varepsilon$ increases with the magnitude $\sigma$ of the
fluctuations.

\begin{figure}[t]
\begin{center}
\begin{tabular}{c}
\includegraphics[width=13cm]{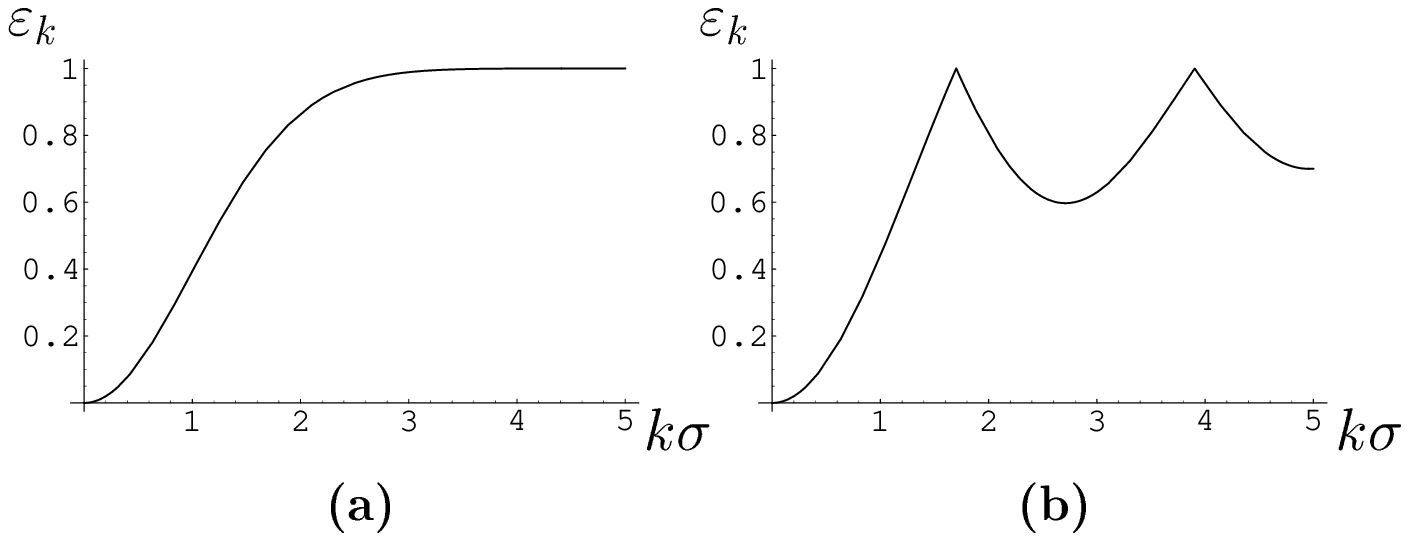}
\end{tabular}
\end{center}
\caption{(a) Decoherence parameter
 $\varepsilon_k$ (\ref{eq:rum4}) versus $k\sigma$,
 for a monochromatic beam interacting with a
 shifter fluctuating according to (\ref{eq:rum1});
 (b) Decoherence parameter
 $\varepsilon_k$ (\ref{eq:dec2}) versus $k\sigma$,
 for a monochromatic beam interacting with
 a shifter fluctuating according to (\ref{eq:sindis}).}
\label{fig:paga}
\end{figure}

For a monochromatic beam [$P_{\rm in}(p')=\delta(p-p')$]
\andy{rum4}
\beq\label{eq:rum4}
\varepsilon_k=1- e^{-\frac{k^2\sigma^2}{2}},
\eeq
with $k=p/\hbar$. This is shown in Figure \ref{fig:paga}(a) and
can be obtained from (\ref{eq:rum6}) in the $\delta \to \infty$
limit. Notice that high momenta are more fragile against
fluctuations \cite{RS}. Moreover, when the distribution of the
shifts is Gaussian, $\varepsilon_k$ and equivalently ${\cal V}(p)$
are monotonic functions: they both depend ``smoothly'' on
$\sigma$.

Let now the phase shifts be distributed according to the law
\cite{fibonacci}
\andy{sindis}
\beq\label{eq:sindis}
w(\Delta-\Delta_0)=\frac{1}{\pi}\frac{1}{\sqrt{2\sigma^2-
(\Delta-\Delta_0)^2} },
\eeq
for $|\Delta-\Delta_0|\leq \sqrt 2 \sigma$ and $0$ otherwise, with
standard deviation $\left(\int \Delta^2 w(\Delta)
d\Delta\right)^{1/2} = \sigma$. From an experimental perspective
this is more convenient and easier to reproduce than the Gaussian
distribution (\ref{eq:rum1}): indeed, (\ref{eq:sindis}) follows
from a phase $\Delta(t)=\Delta_0 + \sqrt 2\sigma\sin t$, where $t$
(``time") is a parameter, uniformly distributed between 0 and
$2\pi$, namely $w(\Delta)=\int_0^{2\pi} dt\;\delta(\Delta-\sqrt 2
\sigma\sin t) /2\pi$. (One can require $\sqrt 2 \sigma
\leq \Delta_0$, in order that $\Delta(t)$ be positive---and the term
$\sqrt 2 \sigma\sin t$ be regarded as a ``small" fluctuation
around the average value. However, strictly speaking, this is not
necessary from a mathematical point of view.) From
(\ref{eq:sindis}) and (\ref{eq:wtilde}) one gets
\andy{lewa}
\barr
\Omega(p) &=&\int_{-\sqrt 2\sigma }^{\sqrt 2\sigma}\;
\frac{d\Delta}{\pi}\frac{e^{i\frac{p\Delta}{\hbar}}}
{\sqrt{2\sigma^2-\Delta^2} } \nonumber \\
& =&\int_{-\pi/2}^{\pi/2 }\frac{dt}{\pi}\exp\left(i\frac{\sqrt
2p\sigma}{\hbar}\sin t\right)= J_0\left(\frac{\sqrt
2p\sigma}{\hbar}\right),
\label{eq:lewa}
\earr
where $J_0$ is the Bessel function of order zero. The decoherence
parameter (\ref{eq:decimp}) reads
\andy{dec1}
\beq\label{eq:dec1}
\varepsilon=1-\max_{\Delta_0}\left|\int\;dp\;P_{\rm in}(p)
J_0\left(\frac{\sqrt 2
p\sigma}{\hbar}\right)\cos\left(\frac{p\Delta_0}
{\hbar}\right)\right|
\eeq
and for a monochromatic beam one obtains ($k=p/\hbar$)
\andy{dec2}
\beq\label{eq:dec2}
\varepsilon_{k}=1-\max_{\Delta_0}\left|J_0\left(\frac{\sqrt 2p\sigma}
{\hbar}\right)
\cos\left(\frac{p\Delta_0}{\hbar}\right)\right|=1-|J_0(\sqrt 2k\sigma)|.
\eeq
This function is shown in Figure \ref{fig:paga}(b): observe that
decoherence is {\em not} a monotonic function of the noise
$\sigma$ in (\ref{eq:sindis}).

A comparison between Figures \ref{fig:paga}(a) and
\ref{fig:paga}(b) is interesting. In both cases one observes
fragility at high momenta $p=\hbar k$. However, the behavior of
decoherence in Figure \ref{fig:paga}(b) is somewhat anomalous and
against naive expectation. For a given $k$, there are situations
where decoherence $\varepsilon$ {\em decreases by increasing} the
strength of the fluctuations $\sigma$. Note also that we are
considering incoming monochromatic beams, whence, according to
(\ref{eq:qwe}) and (\ref{eq:decimp}), $\epsilon_k=1-{\cal V}(\hbar
k)$ and the decoherence parameter is strictly related to the
standard visibility of the interference pattern. Therefore, in the
anomalous regions, one observes an increase in visibility by
increasing the fluctuations of the phase shifter, a phenomenon
somewhat similar to stochastic resonance \cite{stochastic}. This
is true not only for monochromatic beams, but also for narrow
distributions (packets) in momentum space.

These anomalous results are not entirely surprising, if one
compares them to other known results in classical optics. We will
therefore recall in the next section some notions related to the
visibility of a classical interference experiment: the visibility
can be expressed as the Fourier transform of the spectral
distribution of a quasi-monochromatic light source and it displays
some ``anomalies'' even in cases that are different from our
``Gaussian" example (\ref{eq:rum5}).

%%%%%%%%%%%%%%%%%%%%%%%%%%%%%%%%%%%%%%%%%%%%%%%%%%%%%%%%%%%%%%%%%
%%%%%%%%%%%%%%%%%%%%%%%%%%%%%%%%%%%%%%%%%%%%%%%%%%%%%%%%%%%%%%%%%
%%%%%%%%%%%%%%%%%%%%%%%%%%%%%%%%%%%%%%%%%%%%%%%%%%%%%%%%%%%%%%%%%

\section{A classical analogy}
\label{sec-class}
\andy{sec-class}

The phenomena analyzed in the previous sections have an
interesting classical counterpart that is worth looking at in some
detail. In this section we will examine the behavior of the
visibility in a two-beam interference experiment, in relation to
the spectral density distribution of the source. We follow Born
and Wolf \cite{BornW}. Suppose to have two beams whose optical
difference is $\Delta {\cal S}$ and whose wave number is $k=2\pi
/\lambda$. Their phase difference reads
\andy{phdiff}
\beq \label{eq:phdiff}
{\cal \delta}(k,\Delta{\cal S})=k \Delta{\cal S},
\eeq
and, assuming that they have the same intensity ${\sf i}(k)dk$ in
the range $[k, k+dk]$, the intensity at the screen due to the
elementary wave number range $dk$ reads
\andy{idkzero}
\beq \label{eq:idkzero}
i(k,\Delta{\cal S})dk=2{\sf i}(k)[1+\cos (k \Delta{\cal S})] dk.
\eeq
Observe that the different spectral components add incoherently,
so
\andy{intidk}
\beq \label{eq:intidk}
I(\Delta {\cal S})=2\int dk\;{\sf i}(k)[1+\cos (k\Delta {\cal S})]
\eeq
is the intensity at the screen as a function of $\Delta {\cal S}$,
due to both interfering beams. The quantity $k$ is to be compared
to the phase $\Delta$ in Sec.\ \ref{sec-nof}.

In some cases one deals with light sources that emit with
characteristic spectral lines. If we consider only one of these
spectral lines, ${\sf i}(k)$ is different from zero only in a very
small range of $k$ about some mean value $k_0$. Putting
\andy{defx,defj}
\barr \label{eq:defx}
j(k)=2{\sf i}(k_0 + k),
\label{eq:defj}
\earr the intensity at the screen (\ref{eq:intidk}) becomes
\andy{intjdx}
\barr \label{eq:intjdx}
I(\Delta {\cal S})&=&\int dk\; j(k)\{1+\cos [(k_0 +k)\Delta {\cal S}]\} \nonumber \\
&=&N[1+C(\Delta {\cal S})\cos (k_0 \Delta {\cal S})- S(\Delta
{\cal S}) \sin (k_0\Delta {\cal S})], \earr where $N$ is a
normalization factor, defined as the sum of both the (equal)
intensities of the beams, and $C$ and $S$ are the average value on
the spectral distribution $j(k)$ of $\cos(k \Delta {\cal S})$ and
$\sin (k \Delta {\cal S})$ respectively
\andy{defN,defC,defS}
\barr
\label{eq:defN}
& & N=\int dk\;j(k), \\
\label{eq:defC}
& & C(\Delta {\cal S})=\frac{1}{N}\int dk\;j(k)\cos (k \Delta {\cal S}), \\
\label{eq:defS}
& & S(\Delta {\cal S})=\frac{1}{N}\int  dk\;j(k)\sin (k \Delta
{\cal S}), \earr i.e. $C$ and $S$ are respectively the real and
the imaginary part of the Fourier transform $\Omega(\Delta {\cal
S})$ of $j(k)/N$
\andy{realC,imagS}
\barr
C(\Delta {\cal S})={\rm Re}\;\Omega(\Delta {\cal S}), \qquad
S(\Delta {\cal S})={\rm Im}\;\Omega(\Delta {\cal S}), \label{eq:realC} \\
\Omega(\Delta {\cal S})=\int dk \;\frac{j(k)}{N}\;e^{i k \Delta
{\cal S}}.
\label{eq:imagS}
\earr
{}From (\ref{eq:intjdx}), the intensity at the screen can be
written as
\andy{Iend}
\beq \label{eq:Iend}
I=N\left[1+|\Omega(\Delta {\cal S})|\cos(k_0 \Delta {\cal
S}+\varphi(\Delta {\cal S}))\right],
\eeq
where $\tan \varphi (\Delta {\cal S})= S(\Delta {\cal S})/C(\Delta
{\cal S})$. A comparison with Eq.\ (\ref{eq:imptr}) shows that
$\Delta {\cal S}$ plays the same role of $p$.

Because $j(k)$ is very peaked about $k=0$, variations of $C$ and
$S$ can be considered negligible compared with $\cos (k_0 \Delta
{\cal S})$ and $\sin(k_0 \Delta {\cal S})$ in Eq.\
(\ref{eq:intjdx}); analogously for $\varphi$ in (\ref{eq:Iend}).
Consequently, under the assumption of slowly varying envelope, one
can define a ``local'' visibility
\cite{BornW}, given by
\andy{defvis}
\beq
\label{eq:defvis}
{\cal V}(\Delta {\cal S})=\frac{I(\Delta {\cal S})_{\rm MAX} -
I(\Delta {\cal S})_{\rm min}} {I(\Delta {\cal S})_{\rm MAX} +
I(\Delta {\cal S})_{\rm min}}=\left|\Omega(\Delta {\cal
S})\right|,
\eeq
expressed as a function of the optical path difference $\Delta
{\cal S}$. The visibility is therefore the amplitude of the cosine
function in Eq.\ (\ref{eq:Iend}). Observe that, whenever $j(k)$ is
an even spectral distribution,
\andy{evensp}
\beq\label{eq:evensp}
{\cal V}(\Delta {\cal S})=|C(\Delta {\cal S})|=|\Omega(\Delta
{\cal S})|
\eeq
and it is possible to determine (apart from the sign) the Fourier
transform of $j(k)/N$ from the visibility.

\begin{figure}[t]
\begin{center}
\begin{tabular}{c}
\includegraphics[width=14cm]{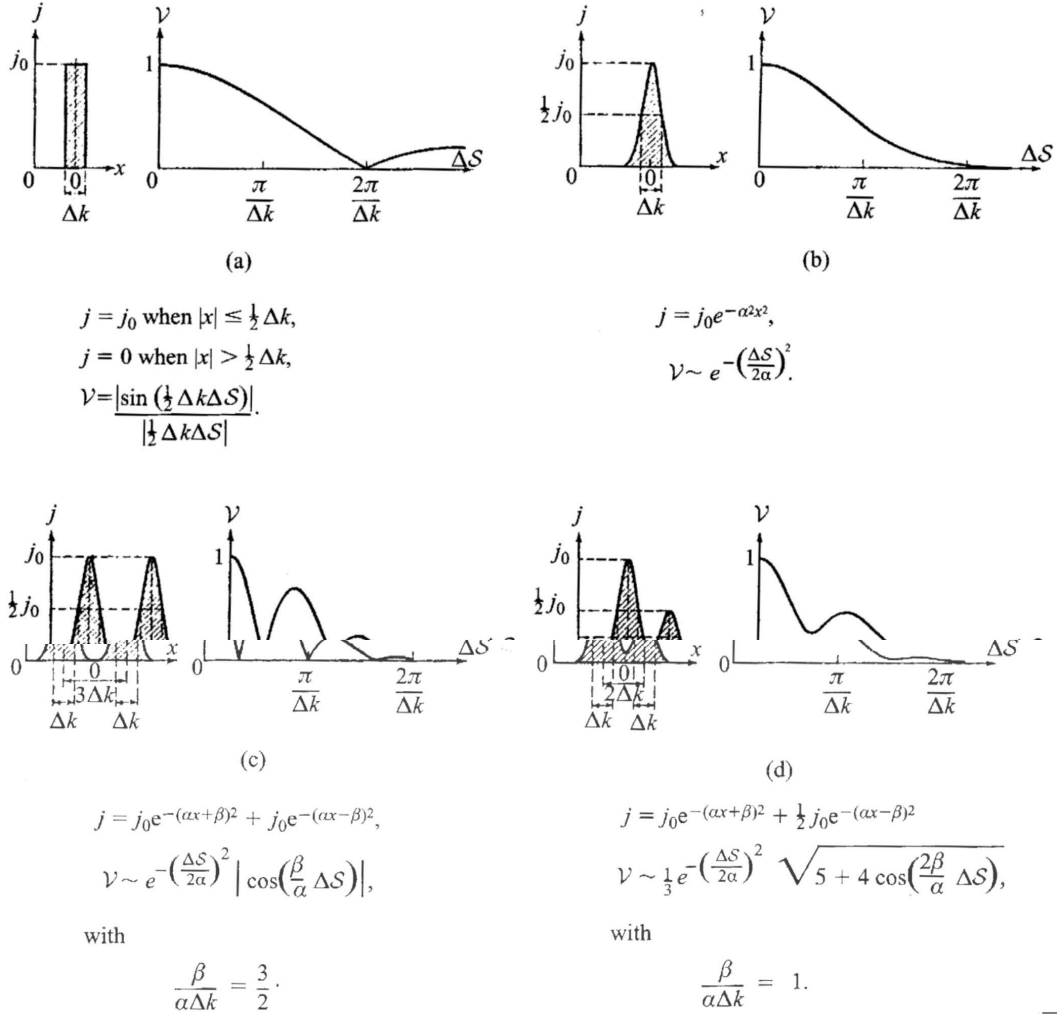}
\end{tabular}
\end{center}
\caption{Visibility versus optical path difference $\Delta {\cal
S}$, for a quasi-monochromatic light source with spectral
distribution $j(x)$ (\ref{eq:defj}). Different spectral
distribution shapes are considered: square-like (a), Gaussian (b),
double Gaussian with peaks at the same level (c) and at different
levels (d). In (b), (c) and (d) $\Delta k=2\sqrt{\ln 2}/\alpha$.
(Reproduced with permission from Ref.~\protect\cite{BornW}.) }
\label{fig:bornwolf}
\end{figure}

Equations (\ref{eq:visibp}) and (\ref{eq:defvis}) are easily
compared. The distribution of the phase shifts in the quantum case
is replaced by the spectral distribution of the incoherent light
source in the classical case. Indeed, $\Omega(\Delta{\cal S})$,
$C(\Delta{\cal S})$ and $S(\Delta {\cal S})$ in
(\ref{eq:realC})-(\ref{eq:imagS}) correspond to $\Omega(p)$, $C(
p)$ and $S( p)$ in (\ref{eq:wtilde})-(\ref{eq:imome}), i.e. the
Fourier transform of $j(k)/N$ corresponds to that of $w(\Delta)$
(notice that $w(\Delta)$ is normalized to unity).

The visibility curves (\ref{eq:evensp}) are shown in
Fig.~\ref{fig:bornwolf} for different shapes of the spectral
distribution $j(k)$. As one can see, they show different behavior.
In Fig.~\ref{fig:bornwolf}(a) a square-like spectral distribution
gives rise to a visibility function $|\sin y/y|$, in Fig.\
\ref{fig:bornwolf}(b) a Gaussian spectral distribution produces a
Gaussian visibility function, in Fig.\ \ref{fig:bornwolf}(c) and
(d) two ``double Gaussian'' distributions (with the peaks that
have or do not have the same level, respectively) yield more
complicate visibility functions. Only in the case (b), i.e. with a
Gaussian spectral distribution, the visibility is a monotonic
function of the optical path difference $\Delta {\cal S}$. In such
a case, the naive expectation is confirmed that, by increasing the
optical path difference, the visibility decreases. This is not
true in cases (a), (c) and (d), where the visibility is not a
decreasing function for every range of $\Delta{\cal S}$, but there
are regions on the screen where, by increasing the optical path
difference, the two-beam interference visibility increases.

Similar results can be obtained if one considers two-beam
interference with extended monochromatic light sources. In such a
case, the source is treated as a collection of monochromatic
point-like sources that add incoherently and, instead of $j(k)dk$,
one deals with ${\sf i}(\alpha)d\alpha$, the elementary intensity
due to such point-like sources of angular width $d\alpha$. As a
result, the visibility is related to the normalized Fourier
transform of the extended source angular intensity distribution.
This problem was already studied at the end of the 19th century
\cite{Fiz19th} and led Michelson to the construction of his
stellar interferometer \cite{stellar}.

\section{Wigner function in the ordinary channel}
\label{sec-wig}
\andy{sec-wig}

In the previous sections we have proposed a definition of
decoherence based on the visibility of the quantum interference
pattern. As we have seen, this definition has some unexpected
features, somewhat at variance with expectation. We also found an
analogy in classical optics. However, alternative definitions of
decoherence are possible, based on the density matrix and on the
Wigner function. Let us therefore briefly recall the definition
and some properties of the Wigner function.

The Wigner quasidistribution function \cite{Wigner} can be defined
in terms of the density matrix $\rho$ as
\andy{wigrho}
\beq\label{eq:wigrho}
W(x,k) = \frac{1}{2\pi}\int d\xi\; e^{-ik\xi} \langle
x+\xi/2|\rho|x-\xi/2\rangle ,
\eeq
where $x$ and $p=\hbar k$ are the position and momentum of the
particle. One easily checks that the Wigner function is normalized
to unity and its marginals represent the position and momentum
distributions
\andy{normW,margx,p}
\barr & & {\rm Tr}\rho= \int dx\; dk\; W(x,k) =1,
\label{eq:normW}\\
& & P(x) = \bra{x}\rho\ket{x}= \int dk\; W(x,k), \label{eq:margx} \\
& & P(k) = \bra{k}\rho\ket{k} =\int dx\; W(x,k).
\label{eq:margp}
\earr
The analyses of the properties of quantum states based on the
Wigner function are useful because they enable one to make prompt
comparisons with fields like quantum optics \cite{QOpt} and
quantum tomography \cite{qtom}.

We focus on one-dimensional systems and assume that the wave
function is well approximated by a Gaussian
\andy{gauss,gaussinv}
\barr \psi(x) &=& \bra{x}\psi\rangle=
\frac{1}{(2\pi\delta^2)^{1/4}} \exp
\left[-\frac{(x-x_0)^2}{4\delta^2} + i k_0 x\right] ,
\label{eq:gauss} \\
\phi(k) &=& \bra{k}\psi\rangle=\frac{1}{(2\pi\delta_k^2)^{1/4}}
\exp \left[-\frac{(k-k_0)^2}{4\delta_k^2} - i(k-k_0)x_0\right]
\nonumber \\
 &=&
\left(\frac{2\delta^2}{\pi} \right)^{1/4} \exp
\left[-\delta^2(k-k_0)^2 - i(k-k_0)x_0\right],
\label{eq:gaussinv}
\earr
where $\psi(x)$ and $\phi(k)$ are the wave functions in the
position and momentum representation, respectively, $\delta$ is
the spatial spread of the wave packet, $\delta_k \delta= 1/2$,
$x_0$ is the initial average position of the particle and
$p_0=\hbar k_0$ its average momentum. The two functions above are
both normalized to one. The Wigner function for the state
(\ref{eq:gauss})-(\ref{eq:gaussinv}) is readily calculated
\andy{wiggauss}
\beq\label{eq:wiggauss}
W(x,k) = \frac{1}{\pi} \exp
\left[-\frac{(x-x_0)^2}{2\delta^2}\right] \exp
\left[-2\delta^2(k-k_0)^2\right].
\eeq
Consider now a neutron wave packet that is split and then
recombined in an interferometer, with a phase shifter $\Delta$
placed in one of the two routes. The Wigner function in the
ordinary channel (transmitted component) is readily computed:
\andy{intcats}
\barr W_{\rm
O}(x,k,\Delta)&=&\frac{1}{4\pi}\exp[-2\delta^2(k-k_0)^2]
\left[\exp\left(-\frac{\left(x-x_0+\Delta\right)^2}{2\delta^2}\right)
\right. \nonumber\\ & & \left.
+\exp\left(-\frac{\left(x-x_0\right)^2}{2\delta^2}\right) \right.
\nonumber\\ & & \left.
+2\exp\left(-\frac{\left(x-x_0+\frac{\Delta}{2}\right)^2}
{2\delta^2}\right)\cos(k\Delta)\right].
\label{eq:intcats} \earr Notice that, for
$\Delta\neq0$, it is not normalized to unity (some neutrons end up
in the extraordinary channel---reflected component) and that for
$\Delta=0$ (no phase shifter) one recovers (\ref{eq:wiggauss}).

\section{Alternative definition of decoherence}
\label{sec-altdec}
\andy{sec-altdec}

We look at a particular case and assume that the shifts $\Delta$
fluctuate around their average $\Delta_0$ according to the
Gaussian law (\ref{eq:rum1}). The average Wigner function reads
\andy{wigmc}
\beq\label{eq:wigmc}
\overline{W}(x,k)= \int d\Delta\; w(\Delta)\; W(x,k,\Delta)
\eeq
and represents a partially mixed state. Essentially, this Wigner
function represents the whole ensemble of neutrons in an
experimental run. For the double Gaussian state
(\ref{eq:intcats}), obtained when a neutron beam crosses an
interferometer, the average Wigner function in the ordinary
channel reads
\andy{catsint}
\barr & &{\overline W}_{\rm O}(x,k)=
\frac{\exp[-2\delta^2(k-k_0)^2]}{4\pi}
 \nonumber\\& &
\times\left\{\exp\left[-\frac{x^2} {2\delta^2} \right]
+\sqrt{\frac{\delta ^2}{\delta^2 + \sigma^2}}
\exp\left[-\frac{(x+\Delta_0)^2} {2(\delta^2+\sigma^2)}\right]
 \right.\nonumber\\& &\left.
+2\sqrt{\frac{\delta ^2} {\delta^2+\frac{\sigma^2}{4}}}
\exp\left[-\frac{\left(x+\frac{\Delta_0}{2}\right)^2+k^2\delta^2
\sigma^2} {2\left(\delta^2+\frac{\sigma^2}{4}\right)}\right]
\cos\left(k\frac{2\delta^2\Delta_0-x\sigma^2}
{2\left(\delta^2+\frac{\sigma^2}{4}\right)}\right)\right\},
\label{eq:catsint}
\earr where we set $x_0=0$ for simplicity. Its momentum marginal
(\ref{eq:margp}) (momentum distribution function) can be computed
analytically and is of interest, because it displays fragility at
high momenta \cite{RS,BRSW}:
\andy{margwn}
\beq\label{eq:margwn}
P(k)=\sqrt{\frac{\delta^2}{2\pi}}\exp\left[-2\delta^2(k-k_0)^2\right]
\left[1+\exp\left(-\frac{k^2\sigma^2}{2}\right)\cos(k\Delta_0)\right].
\eeq

\begin{figure}
\begin{center}
\begin{tabular}{c}
\includegraphics[width=15cm]{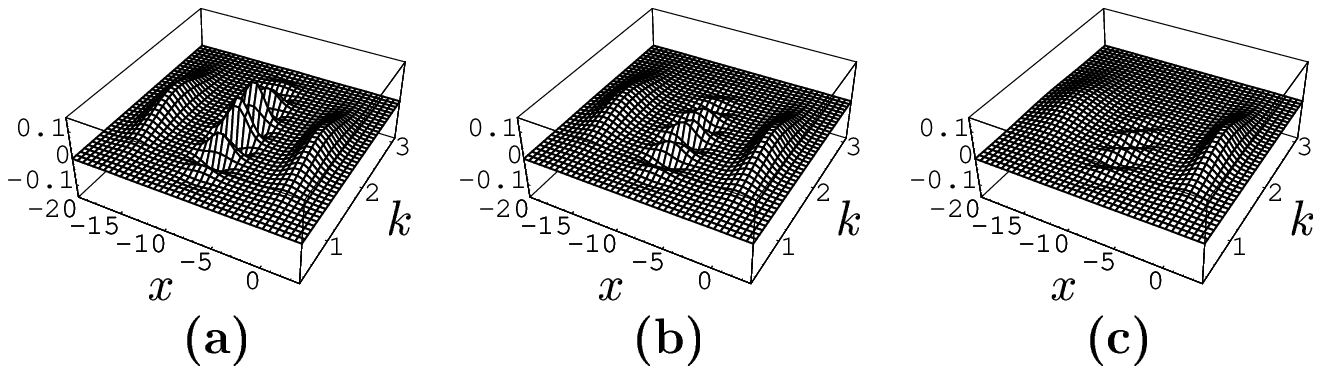}
\end{tabular}
\end{center}
\caption{Wigner function in the ordinary channel
(\ref{eq:catsint}) for different values of the standard deviation
$\sigma$ in (\ref{eq:rum1}). From left to right, $\sigma=0$,
$0.9$, $1.8$\AA. We set $x_0=0$, $k_0=1.7$\AA$^{-1}$,
$\delta=1.1$\AA, $\Delta_0=16.1$\AA. Position $x$ and momentum $k$
are measured in \AA\ and \AA$^{-1}$, respectively. Notice the
strong suppression of interference at large values of momentum,
both in (b) and (c). The interference term in (\ref{eq:catsint})
depends on $x$ and the oscillating part of the Wigner function is
bent towards the negative $x$-axis.}
\label{fig:supp}
\end{figure}

The average Wigner function (\ref{eq:catsint}) is shown in Figure
\ref{fig:supp}. One clearly observes a strong (exponential)
suppression of interference at high values of $k$. Notice that
the oscillating part of the Wigner function is bent towards the
negative $x$-axis. This is due to the $x$-dependence of the cosine
term in (\ref{eq:catsint}) that entails different frequencies for
different values of $x$.

The loss of quantum coherence is clearly visible in Figure
\ref{fig:supp} as the level of noise $\sigma$ increases. One can
try to corroborate this qualitative conclusion by introducing a
quantitative notion of decoherence based on the Wigner function;
however, as we shall see in a while, one runs into the same kind
of difficulties encountered in Sec.\ \ref{sec-capa}. We first
recall that there is an interesting relation between the square of
the Wigner function and the square of the density matrix:
\andy{tracesq}
\beq\label{eq:tracesq}
\int dx\;dk\;W(x,k)^2=\frac{{\rm Tr}\rho^2}{2\pi}.
\eeq
It is therefore possible to define an alternative decoherence
parameter
\cite{fibonacci}, that takes into account the coherence properties of
the neutron ensemble
\andy{decpar}
\beq\label{eq:decpar}
\varepsilon =1-\frac{\mbox{Tr}\rho^2}{(\mbox{Tr}\rho)^2} = 1 -
\frac{2\pi \int dx\; dk\; \overline W (x,k)^2} {\left(\int dx\; dk\;
\overline W(x,k)\right)^2} .
\eeq
This quantity measures the degree of ``purity" of a quantum state:
it is maximum when the state is maximally mixed
($\mathrm{Tr}\rho^2<\mathrm{Tr}\rho$) and vanishes when the state
is pure ($\mathrm{Tr}\rho^2=\mathrm{Tr}\rho$): in the former case
the fluctuations of $\Delta$ are large and the quantum mechanical
coherence is completely lost, while in the latter case $\Delta$
does not fluctuate and the quantum mechanical coherence is
perfectly preserved. The parameter (\ref{eq:decpar}) was
introduced within the framework of the so-called ``many Hilbert
space" theory of quantum measurements \cite{NPN} and yields a
quantitative estimate of decoherence. The related quantity
$\mathrm{Tr}\rho-\mathrm{Tr}\rho^2$ was first considered by
Watanabe \cite{Watanabe} in 1939 (!). A quantity related to
$\varepsilon$ was also introduced in order to get a quantitative
estimate of information for a quantum system \cite{Bru}.

It is also worth noticing that the notion of decoherence just
introduced is based on the square of the density matrix (or Wigner
function) and therefore is not accessible to a direct measurement
procedure. In this sense, it is less ``operational" than that
discussed in Sec.\ \ref{sec-cont}.

\begin{figure}
\begin{center}
\begin{tabular}{c}
\includegraphics[width=8.5cm]{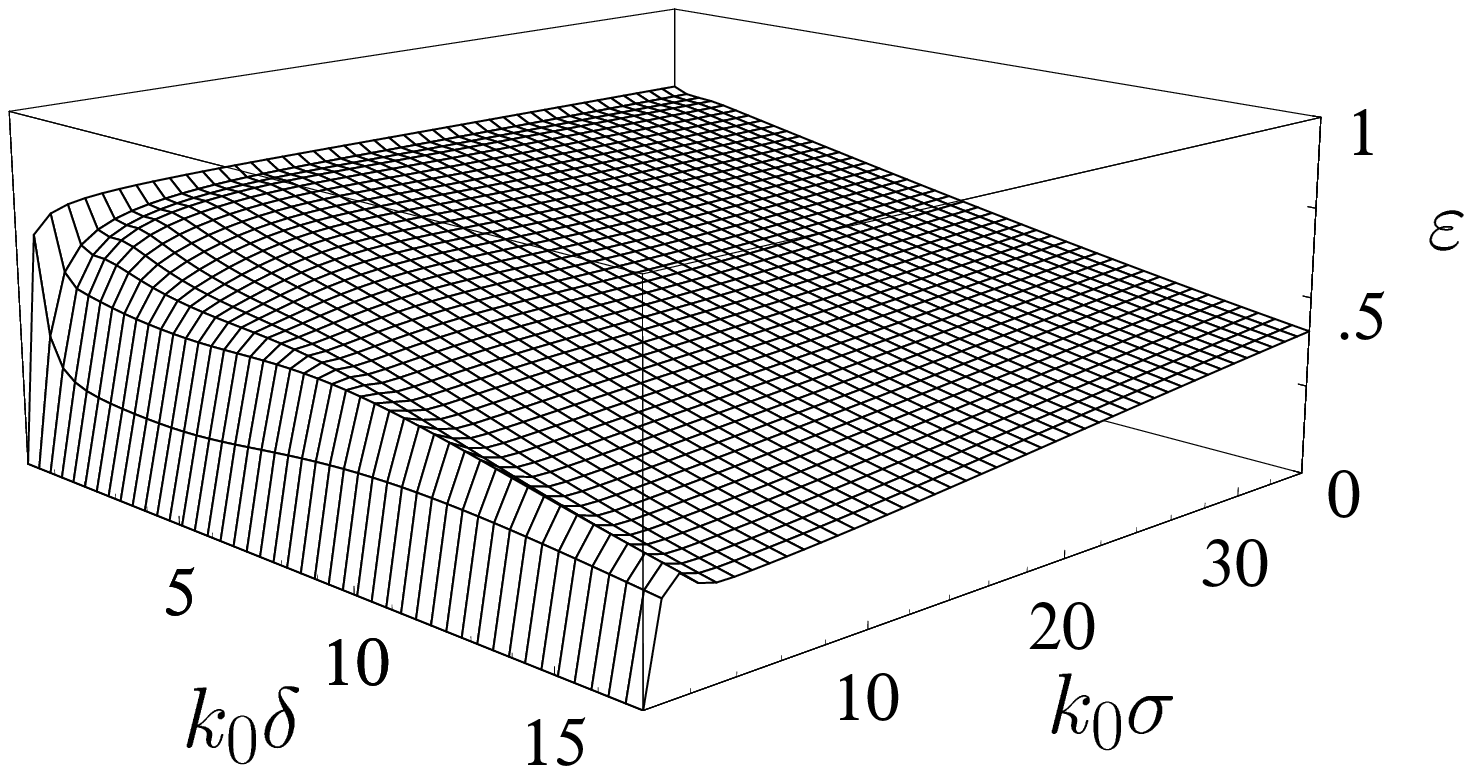}
\end{tabular}
\end{center}
\caption{Decoherence parameter vs coherence length of the wave
packet $\delta$ and standard deviation of the fluctuation $\sigma$
for a double Gaussian wave packet (\ref{eq:intcats}) in the
ordinary channel of a MZI. We set $k_0\Delta_0=27.4$. The
decoherence parameter is not a monotonic function of $\sigma$ for
every value of $\delta$. Notice that $\varepsilon$ never reaches
unity ($\varepsilon\leq 3/4$): this is due to the fact that only
one Gaussian (in one branch of the interferometer) undergoes
statistical fluctuations (see Eq.\ (\ref{eq:catsint}) and Figure
\ref{fig:supp}).}
\label{fig:decabc}
\end{figure}

The decoherence parameter (\ref{eq:decpar}) is shown in Figure
\ref{fig:decabc} as a function of the coherence length of
the wave packet $\delta$ in (\ref{eq:gauss})-(\ref{eq:intcats})
and the standard deviation of the fluctuations $\sigma$. It is
{\em not} a monotonic function of $\sigma$ for all values of
$\delta$. Once again, like in Sec.\ \ref{sec-capa}, there are
situations in which a larger noise yields a more coherent wave
packet (according to a given definition).
The behavior of $\varepsilon$ has a nontrivial dependence both on
the fluctuations ($\sigma$) and on the wave packet properties
($k_0$ and $\delta$).

\section{Entropy}
\label{sec-entr}
\andy{sec-entr}

The conclusions of the previous sections can be corroborated and
put on a somewhat sounder basis by computing the entropy of the
distribution of the shifts according to the formula
\andy{entr}
\beq\label{eq:entr}
S= -\int d\Delta\; w(\Delta)\; \log w(\Delta).
\eeq
This quantity yields an estimate of the collective ``degree of
disorder" of the distribution of the shifts $w(\Delta)$. One can
draw general conclusions about the behavior of $S$ as a function
of a parameter $\sigma$ characterizing the width of the
distribution. Indeed, let $w(\Delta;\sigma)$ be the symmetric
distribution with the properties (\ref{eq:wrules}), $\sigma$ being
its standard deviation. By assuming that the distribution function
$w$ depends only on the single dimensional parameter $\sigma$,
then it must scale according to
\andy{dim}
\beq\label{eq:dim}
w(\Delta;\sigma) =
\frac{1}{\sigma}w\left(\frac{\Delta}{\sigma};1\right).
\eeq
Therefore
\andy{entrsigma}
\barr
S (\sigma)&=& -\int d\Delta\; w(\Delta;\sigma)\; \log
w(\Delta;\sigma)
\nonumber \\
&=& -\int \frac{d\Delta}{\sigma}\;
w\left(\frac{\Delta}{\sigma};1\right) \; \log\left[
\frac{1}{\sigma} w\left(\frac{\Delta}{\sigma};1\right)\right]
\nonumber \\
&=& -\int d\Delta'\; w(\Delta';1)\; \log w(\Delta';1)+ \int
d\Delta'\; w(\Delta';1)\; \log \sigma
\nonumber \\
&=& S(1) + \log \sigma,
\label{eq:entrsigma}
\earr
where $S(1)$ is independent of $\sigma$ and depends only on the
form of the distribution function. $S(\sigma)$ is clearly an
increasing function of $\sigma$.

For example, the Gaussian distribution (\ref{eq:rum1}) yields
\cite{Suda}
\andy{entwn}
\beq\label{eq:entwn}
S(\sigma)=\log\sigma + \frac{1}{2}\log(2\pi e),
\eeq
while the ``sine" distribution (\ref{eq:sindis}) yields
\andy{entwn1}
\beq\label{eq:entwn1}
S(\sigma)=\log\sigma - \frac{1}{2}\log 2.
\eeq
Therefore, the behavior of the decoherence parameter $\varepsilon$
as a function of the entropy $S$ of the shifts is qualitatively
equivalent to its behavior as a function of the standard deviation
$\sigma$. Indeed, Figs.\ \ref{fig:gaun}, \ref{fig:paga} and
\ref{fig:decabc} would differ only for a logarithmic scale on the
abscissae. As we have seen in this article, in general, the two
quantities $S$ and $\varepsilon$ do not necessarily agree: in
other words, the loss of quantum mechanical coherence is not
necessarily larger when the neutron beam interacts with
fluctuating shifts of larger entropy.

\section{Conclusions}

We have introduced and discussed some interference experiments
that display some ``anomalies" both in the classical and in the
quantum domains. The neutron beam partially looses its quantum
coherence as a consequence of the fluctuations of the phase shifts
$\Delta$. One should emphasize that we have considered the case of
``slow" fluctuations, in the sense that each neutron crosses a
phase shifter of length $L$, but the length of the shifter varies
for different neutrons in the beam (different ``events"). We have
supposed that every neutron undergoes a shift $\Delta$ that is
statistically distributed according to a distribution law
$w(\Delta)$.

We focussed our attention on two alternative decoherence
parameters. The first is defined in terms of a generalized
visibility of the interference pattern in a double-slit experiment
(MZI) and is more operational. The second hinges upon less
operational concepts, such as the square of the density matrix.

All our results corroborate the ideas expressed elsewhere
\cite{fibonacci} and make it apparent that the concept of loss
of quantum mechanical coherence deserves clarification and
additional investigation. It would also be interesting to discuss
analogies and differences with conceptual experiments in which
decoherence is complemented by Welcher-Weg information
\cite{englert}.

\ack We wish to thank Prof.\ E.\ Wolf, who kindly granted us
permission to reproduce Fig.~7.54 from Ref.\ \cite{BornW}.

\section*{References}

%%%%%%%%%%%%%%%%%%%%%%%%%%%%%%%%%%%%%%%%%%%%%%%%%%%%

\end{document}